\documentclass[preprint,aps,floats]{revtex4}
\usepackage{epsfig}

\usepackage{graphicx}
\usepackage{dcolumn}
\usepackage{bm}
\def\lsim{\mathrel {\vcenter {\baselineskip 0pt \kern 0pt
    \hbox{$&lt;$} \kern 0pt \hbox{$\sim$} }}}
\def\gsim{\mathrel {\vcenter {\baselineskip 0pt \kern 0pt
    \hbox{$&gt;$} \kern 0pt \hbox{$\sim$} }}}
\def\beq{\begin{equation}}
\def\eeq{\end{equation}}
\def\bea{\begin{eqnarray}}
\def\eea{\end{eqnarray}}

\begin{document}

\title{Long range forces and limits on unparticle interactions}

\author{N.G. Deshpande, Stephen D.H. Hsu, and Jing Jiang}
\affiliation{ Institute of Theoretical Science, University of Oregon,
Eugene, OR 97403}  

\date{\today}

\bigskip

\begin{abstract}
Couplings between standard model particles and unparticles
from a nontrivial scale invariant sector can lead to long range
forces. If the forces couple to quantities such as baryon or lepton
(electron) number, stringent limits result from tests of the
gravitational inverse square law. These limits are much stronger than
from collider phenomenology and astrophysics.
\end{abstract}


\maketitle

\section{Introduction}
Georgi \cite{Georgi:2007ek,Georgi:2007si} has proposed that
unparticles -- a nontrivial scale invariant sector \cite{Banks:1982gt}
-- might couple to standard model
particles, leading to novel phenomenological signatures. This idea has
been further developed by a number of authors
~\cite{Cheung:2007ue}-\cite{Hamada:2007vu}. In this note, we discuss
the possibility of long range forces resulting from such
interactions. We assume strict conformal invariance down to low
energies, so that the unparticle propagator necessarily has a zero
momentum pole.

Our analysis is quite similar to that of Goldberg and Nath
\cite{Goldberg:2007tt}, who considered the possibility that (exactly
scale invariant) unparticles might couple to the ordinary
energy-momentum tensor $T_{\mu \nu}$. Here, we consider couplings
between unparticles and currents such as $J_\mu = \bar{e} \gamma_\mu
e$ or $\bar{q} \gamma_\mu q$. Interestingly, the electron number
current appears in Georgi's $e^+e^- \rightarrow \mu^+ \mu^-$ example
in~\cite{Georgi:2007si}. We find extremely strong limits on such
couplings, much stronger than can be obtained by collider experiments
and even astrophysics.

\section{Long range force due to vector unparticle}

We consider first baryon current $B_\mu$ in terms of quarks
\beq
B_\mu = \frac{1}{3} (\bar{u} \gamma_\mu u ~+~ \bar{d} \gamma_\mu d ~+~ \cdots)\,.
\eeq
The coupling between a vector unparticle state with quark currents can
be rewritten as
\beq
{\cal L} = \lambda_B \Lambda_U^{1-d_U} B_\mu O_U^\mu\,,
\eeq
Where $\Lambda_U$ is the scale below which operator $O_U$ has
dimension $d_U$ and $\lambda_B$ is a dimensionless coupling constant.
In the static limit this interaction generates the potential
\beq
V_U = \lambda_B^2 \Lambda^{2 - 2 d_U}_U \frac{1}{4 \pi^2}
\frac{1}{r^{2 d_U -1}} A_{d_U} \Gamma(2d_U-2) B_1 B_2\,,
\eeq
where
\beq
A_{d_U} = \frac{16 \pi^{\frac{5}{2}}}{(2 \pi)^{2 d_U}} \frac{\Gamma(d_U +
\frac{1}{2})}{\Gamma(d_U-1) \Gamma(2 d_U)}
\eeq
is the coefficient associated with the transverse four-vector unparticle
propagator~\cite{Georgi:2007si} and $B_{1,2}$ are the baryon numbers
of the two interacting masses.  Using a relationship involving Gamma
functions, $V_U$ can be simplified into
\beq
V_U = \frac{1}{2 \pi^{2 d_U}} \lambda_B^2 \Lambda_U^{2 -2 d_U}
\frac{\Gamma(d_U + \frac{1}{2}) \Gamma(d_U - \frac{1}{2})}{\Gamma(2 d_U)}
\frac{1}{r^{2 d_U -1}} B_1 B_2\,.
\eeq
Note that for $d_U = 1$, this will produce a $1/r$ repulsive potential
\beq
V_U = \frac{1}{4 \pi} \lambda_B^2 \frac{1}{r} B_1 B_2\,.
\eeq

Similarly, the unparticle operator can couple to lepton currents with
coupling strength $\lambda_L$ and the above can be directly applied as
well.  For numerical results, we exhibit three examples: $\lambda_B =
\lambda \ne 0$ and $\lambda_L = 0$; $\lambda_B = 0$ and $\lambda_L =
\lambda \ne 0$; and $\lambda_B = - \lambda_L = \lambda$.  In these
cases, the unparticle operator couples to $B$, $L$ and $B-L$ currents,
respectively.

\section{Numerical results}

In the limit of unbroken scale invariance, the force between $B_1$ and
$B_2$ is a long range force similar to gravity, but it may have different
$1/r$ power dependence.  To obtain numerical results, we make the
approximation $B_{1,2} \approx m_{1,2}/u$, where $m_{1,2}$ are the
masses of the two interacting bodies and $u$ is the atomic mass unit.
Combined with the gravitational potential, the potential between two
objects of mass $m_1$ and $m_2$ is
\beq
V = - G \frac{m_1 m_2}{r} ~+~ \lambda_B^2 f_{d_U} \frac{1}{u^2}
\frac{m_1 m_2}{r^{2 d_U -1}}\,,
\eeq
where Newton's constant $G = 6.7 \times 10^{-39}$ GeV$^{-2}$ and
\beq
f_{d_U} = \frac{1}{2 \pi^{2 d_U}} \Lambda_U^{2 -2 d_U} 
\frac{\Gamma(d_U + \frac{1}{2}) \Gamma(d_U - \frac{1}{2})}{\Gamma(2 d_U)}
\eeq
captures the $d_U$ dependent part of the coefficient.  $V$ can then be
written as
\beq
V = -G \frac{m_1 m_2}{r} ~+~ G \frac{m_1 m_2}{r} \frac{1}{u^2}
\frac{\lambda_B^2}{G} f_{d_U} \left(\frac{1}{r}\right)^{2 d_U -2}\,.
\eeq
We compare the second term on the right hand side to the power-law
potential in Ref.~\cite{Adelberger:2006dh},
\beq
V_{12}^k (r) = - G \frac{m_1 m_2}{r} \beta_k \left(\frac{1~{\rm
mm}}{r}\right)^{k-1}\,,
\label{eq:betak}
\eeq and derive limits on the coupling $\lambda$, for $k=2$ and
$k=3$. Ref.~\cite{Adelberger:2006dh} also discussed constraints on
Yukawa potentials generated by exchange a scalar or a vector boson.
For small enough boson masses $m$, the experimental distance $r$ is
much less than the Compton wavelength $\hbar/mc$.  In this limit, the
Yukawa potential is approximately $1/r$, and we obtain a limit on
the case $k \approx 1$. For non-integer value of $k$, we simply
interpolate the limit linearly (the precise bound will be of the
same order of magnitude).  The range of $d_U$ we are interested
in is $1 \le d_U \le 2$ and it is related to $k$ through equation $2
d_U -2 = k -1$.  The limits on the scalar or vector coupling can be
converted into limits on an effective coefficient $\beta_1$ for the
case of a $1/r$ potential. Note that when we consider the $B$, $L$ and
$B-L$ cases, the effective couplings will be multiplied by factors of
$(Z+N)/A \approx 1$, $Z/A$ and $N/A$ respectively, where $Z$ is the
number of protons, $A$ the atomic number, and $N$ is the number of
neutrons.  For the molybdenum pendulum considered in
Ref.~\cite{Adelberger:2006dh}, $Z/A = 0.438$ and $N/A = 0.563$.  If
$\lambda_B$ and $\lambda_L$ happen to satisfy \beq
\label{special}
\frac{\lambda_B}{\lambda_B + \lambda_L} = - \frac{Z}{N}\,, \eeq the
effective charge becomes zero and the torsion-balance experiment
becomes insensitive to $\lambda$, so no limit is obtained. However,
the experiment \cite{Adelberger:2004} was performed with an aluminum pendulum and copper attractor and arrived at similar bounds, covering as well the exceptional parameter space (\ref{special}).

We should also note that forces coupling to almost any linear combination of $B$ and $L$ which extend over truly macroscopic (e.g., solar system scale) distances are even more tightly constrained if they deviate from $1/r$, since they would affect Newtonian orbits.

Table~\ref{tbl:lambda} shows the $68\%$ confidence level (CL)
constraints on $|\beta_k|$ (first column) and the derived constraints
on $\lambda$ for the case of $B$ current (second column).  The results
for $L$ currents are only different for $d_U = 1$, $|\lambda| < 2.5
\times 10^{-20}$ and for $d_U = 1.25$, $|\lambda| < 5.0 \times
10^{-17}$.  The results for the $B-L$ current are almost identical to that for the $L$ current by accident.  These limits are proportional to
$\Lambda_U^{2-2d_U}$.  The values in the table are for $\Lambda_U = 1$
TeV and limits for other values of $\Lambda_U$ can be easily
calculated by simple scaling.

For comparison purposes, Table~\ref{tbl:lambda} also displays results
from supernova constraints (last column, taking $\Lambda_U = 1$
TeV). Unparticles can induce too-rapid cooling of
supernovae~\cite{Davoudiasl:2007,Hannestad:2007ys}.  A constraint on
$\lambda$ can be deduced, yielding \beq \lambda
\left(\frac{\Lambda_U}{30~{\rm MeV}}\right)^{1-d_U} < 3 \times
10^{-11}\,.  \eeq Note that for our limits to apply scale invariance
must hold to length scales as long as a fraction of a millimeter. If
scale invariance is broken at a scale intermediate between a
millimeter and the thermal wavelength of a supernova (roughly, inverse
MeV), the supernova constraints will still apply, while ours will not.

\begin{table}[htb]
\begin{center}
\begin{tabular}{c|c|c|c}
\hline\hline
$d_U$ & $|\beta_k|$ & $|\lambda|$ & SN\\
\hline
1  & $1.8\times 10^{-2}$ & $3.9 \times 10^{-20}$ & $3.0 \times
10^{-11}$ \\
1.25 & $9.1 \times 10^{-3}$ & $7.5 \times 10^{-17}$ & $4.1 \times
10^{-10}$ \\
1.5  & $4.5 \times 10^{-4}$ & $4.3 \times 10^{-14}$ & $5.5 \times
10^{-9}$ \\
1.75 & $2.9 \times 10^{-4}$ & $8.8 \times 10^{-11}$ & $7.4 \times
10^{-8}$\\
2.0  & $1.3 \times 10^{-4}$ & $1.5 \times 10^{-7}$ & $1.0 \times
10^{-6}$\\
\hline
\end{tabular}
\end{center}
\caption{The $68\%$ CL constraints on the coupling $\lambda$ for the 
$B$ current. SN refers to supernova constraint on $B$ current.}
\label{tbl:lambda}
\end{table}

\section{Conclusions}

A sector of particle physics which exhibits nontrivial scale
invariance would be an exciting discovery. If, however, the scale
invariance is exact, long range forces may result which are already
strongly constrained by measurements of the gravitational inverse
square law.  Given the limits derived here, perhaps the most likely
unparticle scenario is one in which the scale invariance is only
approximate -- i.e., it is broken below some energy scale, thereby
screening the long range forces. However, in this case the new
particle sector, while exhibiting novel dynamics, is not really an
{\it unparticle} sector!

\vskip 1.0cm \noindent {\bf Acknowledgments}$\,$ The authors thank
Dave Soper for discussions, and Eric Adelberger for comments and for clarification regarding the experiments in \cite{Adelberger:2006dh, Adelberger:2004}. This work was supported by Grant
No. DE-FG02-96ER40949 of the U.S.  Department of Energy.

\end{document}